\documentclass[twocolumn,prl,aps,preprintnumbers,longbibliography, superscriptaddress, nobalancelastpage]{revtex4-2}
\usepackage{amsmath,amssymb,bm,graphicx,color,gensymb,bbold,appendix,xspace}
\usepackage[colorlinks=true, citecolor={blue!80!black}, urlcolor={blue!50!black}, linkcolor = {blue!80!black}]{hyperref}
\usepackage[utf8x]{inputenc}

\usepackage{xcolor}

\newcommand{\KCSO}{K$_2$Co(SeO$_3$)$_2$\xspace}

\begin{document}

\title{Transverse order and longitudinal fluctuations in a near-Ising spin supersolid }
\author{Mengze Zhu}
\address{Laboratory for Solid State Physics, ETH Z\"{u}rich, 8093 Z\"{u}rich, Switzerland}

\author{V. Romerio}
\address{Laboratory for Solid State Physics, ETH Z\"{u}rich, 8093 Z\"{u}rich, Switzerland}

\author{S. Raymond}
\address{Université Grenoble Alpes, CEA, IRIG, MEM, MDN, F-38000 Grenoble, France}

\author{N. Murai}
\address{J-PARC Center, Japan Atomic Energy Agency, Tokai, Ibaraki 319-1195, Japan}

\author{S. Ohira-Kawamura}	
\address{J-PARC Center, Japan Atomic Energy Agency, Tokai, Ibaraki 319-1195, Japan}

\author{K.~Yu.~Povarov}
\address{Dresden High Magnetic Field Laboratory (HLD-EMFL) and W\"urzburg-Dresden Cluster of Excellence ctd.qmat, Helmholtz-Zentrum Dresden-Rossendorf, 01328 Dresden, Germany}

\author{S.~A.~Zvyagin}
\address{Dresden High Magnetic Field Laboratory (HLD-EMFL) and W\"urzburg-Dresden Cluster of Excellence ctd.qmat, Helmholtz-Zentrum Dresden-Rossendorf, 01328 Dresden, Germany}

\author{R. Sibille}
\address{PSI Center for Neutron and Muon Sciences, 5232 Villigen PSI, Switzerland}

\author{A. Minelli}
\address{Neutron Scattering Division, Oak Ridge National Laboratory, Oak Ridge, Tennessee 37831, USA}

\author{Z. Yan}
\address{Laboratory for Solid State Physics, ETH Z\"{u}rich, 8093 Z\"{u}rich, Switzerland}

\author{S. Gvasaliya}
\address{Laboratory for Solid State Physics, ETH Z\"{u}rich, 8093 Z\"{u}rich, Switzerland}

\author{A.~Zheludev}
\email{zhelud@ethz.ch; http://www.neutron.ethz.ch/}
\address{Laboratory for Solid State Physics, ETH Z\"{u}rich, 8093 Z\"{u}rich, Switzerland}

\begin{abstract}
We investigate the polarization of the ordered moments and low-energy spin fluctuations in the spin supersolid state of the $S=1/2$ triangular-lattice easy-axis antiferromagnet \KCSO using neutron scattering. The supersolid order develops through successive BKT transitions: the longitudinal order appears at higher temperature, while the transverse component associated with the Bose-Einstein condensate emerges only below a lower-temperature transition accompanied by a change in the interlayer correlations. At the lowest measured temperature, the transverse ordered moment reaches only $\sim$11\% of the longitudinal component. Moreover, the low-energy spin fluctuations are found to be predominantly longitudinal in character. These results provide direct evidence that the supersolid ground state survives even in the near-Ising regime and exhibits longitudinal low-energy dynamics, imposing stringent constraints on microscopic theories of triangular-lattice XXZ antiferromagnets.

\end{abstract}

\date{\today}
\maketitle


The antiferromagnetic $S=1/2$ XXZ model on the triangular lattice is a paradigmatic platform of frustrated magnetism. For easy-axis exchange anisotropy, numerous theoretical studies predict a spin supersolid ground state with a three-sublattice Y spin structure \cite{Miyashita_Kawamura1985,Miyashita1986,Wessel2005,Melko2005,Heidarian2005,Wang2009,Jiang2009,Heidarian2010,Zhang2011,Yamamoto2014,SellmannZhangEggert_PRB_2015_AnotherXXZtriangularPhD,gallegos2025}, where the longitudinal spin component corresponds to a boson density modulation while the transverse component represents a Bose-Einstein condensate (BEC) breaking U(1) symmetry. However, recent exact diagonalization \cite{Ulaga2023,Ulaga2024} and tensor-network calculations \cite{Xu2025} suggest that the BEC order parameter may vanish close to the Ising limit, leaving the existence of supersolidity in the strongly anisotropic regime an unresolved question.

Several recently discovered $S = 1/2$ triangular-lattice antiferromagnets provide promising experimental realizations of this model \cite{Ma2016,MacdougalWilliams_PRB_2020_BaCoSbOtriangularexcitations,Sheng2022,Gao2022,Woodland2025,Arh2022,Scheie2024,Zhong2020,Zhu2024,Zhu2025,Chen2026}. Among them, \KCSO is particularly attractive because its strong easy-axis anisotropy, $J_{xy}/J_z \approx 0.07$, places it near the Ising limit \cite{Zhu2024,Zhu2025,Chen2026}, providing a rare opportunity to answer this question experimentally. Yet direct evidence for transverse order has remained elusive, as the transverse moment is expected to be intrinsically small and is further obscured by the quasi-two-dimensional nature of the magnetic order \cite{Zhu2024,Chen2026}. 

The nature of the low-energy spin excitations in the supersolid state is likewise under active debate \cite{kleine1992,Kleine1992_Perturbationtheory,jia2024,Gao2024,Chi2024,sheng2025,Xu2025,Bose2025,Mauri2026}. Although inelastic neutron scattering (INS) has revealed broad continua and roton-like dispersion minima in the magnetic excitation spectrum of \KCSO \cite{Zhu2024,Zhu2025,Chen2026}, their microscopic origin remains unclear. Recent quantum Monte Carlo (QMC) simulations show that many spectral features can be reproduced by longitudinal fluctuations alone \cite{Zhu2025}, raising the question whether the low-energy excitations of the spin supersolid are primarily longitudinal in character.

Here we address both questions by determining the polarization of the ordered moments and spin fluctuations in \KCSO using neutron scattering. We demonstrate that the transverse ordered moment emerges only below $T\approx0.35$ K, well below the onset temperature of longitudinal order, and reaches only $\sim$11\% of the longitudinal moment at the lowest measured temperature. Despite this small transverse moment, the low-energy spin fluctuations are predominantly longitudinal. These results establish that the supersolid ground state persists in the near-Ising regime and is characterized by weak transverse order and unconventional longitudinal low-energy dynamics, providing stringent experimental benchmarks for microscopic theories of supersolidity in triangular-lattice XXZ antiferromagnets.

\begin{figure}[ht]
	\includegraphics[width=\columnwidth]{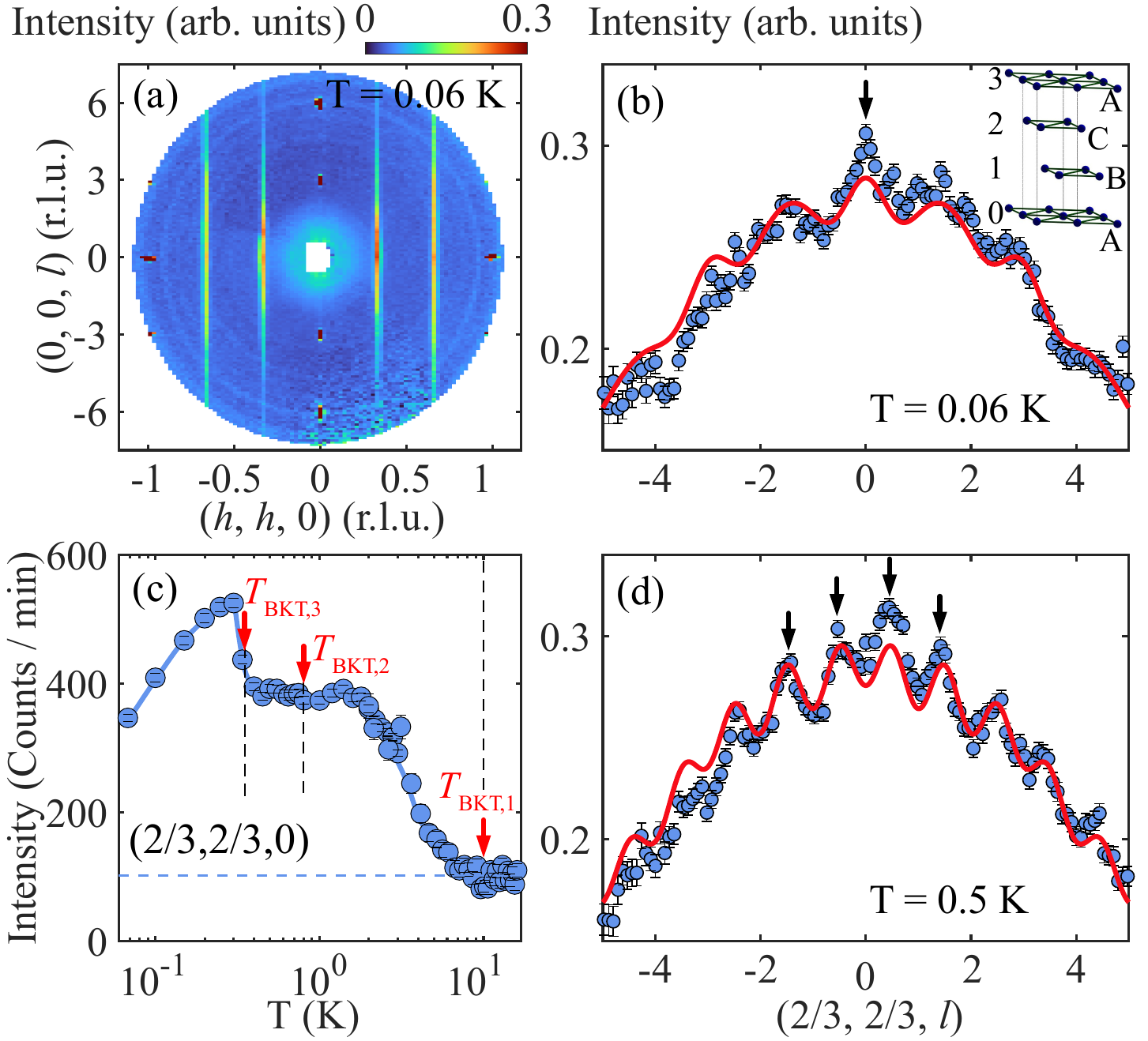}
	\caption{(a) Elastic neutron-scattering intensity of \KCSO in the $(h,h,l)$ plane measured at $T$ = 0.06 K using CORELLI. Rod-like magnetic scattering indicates quasi-two-dimensional magnetic order. (b),(d) $l$ scans at $\mathbf{Q}=(2/3,2/3,l)$ measured at $T$ = 0.06 and 0.5 K, respectively.  Black arrows mark the intensity maxima. Red solid lines are fits described in the text. Inset in (b) illustrates ABC stacking of the triangular-lattice planes. (c) Temperature dependence of the magnetic intensity at $\mathbf{Q}=(2/3,2/3,0)$ measured using ZEBRA. Red arrows mark the BKT transitions identified from specific-heat measurements \cite{Zhu2025}. The enhancement at $T_{\text{BKT,3}}$ coincides with the onset of the transverse ordered moment. The blue dashed line represents high-temperature background. The blue solid line is a guide to the eye. }\label{fig:unpolarized_diffraction}
\end{figure}


Given the distinct energy scales governing longitudinal and transverse ordering, magnetic order is expected to develop in separate steps \cite{Jose77,Sheng1992,Stephan2000,Melchy2009,Heidarian2018}. Previous specific-heat measurements on \KCSO identified three Berezinskii-Kosterlitz-Thouless (BKT) transitions upon cooling, attributed to the successive onset of longitudinal and transverse spin orders. The longitudinal component was proposed to develop through the two higher-temperature transitions at $T_{\text{BKT,1}} \approx 10$ K and $T_{\text{BKT,2}} \approx 0.8$ K, whereas the transverse ordered moment appears only below the lowest-temperature transition at $T_{\text{BKT,3}} \approx 0.35$ K \cite{Zhu2025}. 

To test this scenario, we first performed unpolarized neutron diffraction measurements using the CORELLI diffuse-scattering spectrometer \cite{Ye_Corelli_2018} at Spallation Neutron Source in Oak Ridge National Laboratory and the lifting-counter diffractometer ZEBRA at Paul Scherrer Institute. 
Figure \ref{fig:unpolarized_diffraction}(a) shows the elastic diffraction pattern in the $(h,h,l)$ plane measured at $T$ = 0.06 K. Rod-like magnetic scattering is observed at $\mathbf{Q}=(\pm 1/3,\pm 1/3,l)$ and $(\pm 2/3,\pm 2/3,l)$, consistent with quasi-two-dimensional magnetic order and a three-sublattice spin structure in the triangular plane, in agreement with previous studies \cite{Zhu2024,Chen2026}.
Temperature dependence of the magnetic intensity at $\mathbf{Q} = (2/3,2/3,0)$ is shown in Fig.~\ref{fig:unpolarized_diffraction}(c). Magnetic scattering first appears below $T \approx10$ K, coincident with the highest-temperature BKT transition $T_{\text{BKT,1}}$ \cite{Zhu2025}. Upon further cooling, however, the intensity evolves nonmonotonically, and exhibits a pronounced enhancement near $T_{\text{BKT,3}} \approx$ 0.35 K \cite{Zhu2025}, before decreasing again at lower temperatures.

The intensity enhancement at $T_{\text{BKT,3}}$ is accompanied by a marked change in the interlayer correlations. Representative $l$-scans over $\mathbf{Q} = (2/3,2/3,l)$ at $T = 0.06$ and 0.5 K are shown in Fig.~\ref{fig:unpolarized_diffraction}(b) and \ref{fig:unpolarized_diffraction}(d). At $T$ = 0.5 K, the magnetic intensity is maximized at half-integer $l$, whereas at $T$ = 0.06 K the maximum shifts to the integer position $l$ = 0 (black arrows). 

To directly determine whether a transverse ordered moment emerges at $T_{\text{BKT,3}}$, we performed polarized neutron diffraction measurements using the IN12 cold-neutron diffractometer at the Institut Laue-Langevin on a single crystal aligned in the $(h,k,0)$ scattering plane. 
Polarization analysis was conducted using CRYOPAD \cite{CRYOPAD2005}. The incident neutron energy was $E_i$ = 4.664 meV, with the neutron polarization oriented perpendicular to the scattering plane ($\mathbf{P} \parallel z$). 
In this geometry, the non-spin-flip (NSF) and spin-flip (SF) channels provide complementary sensitivity to the longitudinal and transverse ordered moments. Specifically, for any magnetic reflection $\mathbf{Q}=(h,k,0)$,
\begin{align}
    I_{\text{NSF}} \propto &~\alpha S^{zz}(\mathbf{Q}) + \frac{1-\alpha}{2}S^{\perp}(\mathbf{Q}) + I_{\text{BG,NSF}} \nonumber \\
    I_{\text{SF}} \propto  &~\frac{\alpha}{2} S^{\perp}(\mathbf{Q}) + (1-\alpha)S^{zz}(\mathbf{Q}) + I_{\text{BG,SF}} \nonumber
\end{align}
where $\alpha$ is the fraction of spin-up neutrons in the incident beam, $S^{zz}(\mathbf{Q})$ and $S^{\perp}(\mathbf{Q})= S^{xx}(\mathbf{Q})+S^{yy}(\mathbf{Q})$ denote the longitudinal and transverse static structure factors derived from Fourier transform of the magnetization-magnetization correlation functions, and $I_{\text{BG,NSF}}$ and $I_{\text{BG,SF}}$ represent the background. 

\begin{figure}[t]
	\includegraphics[width=\columnwidth]{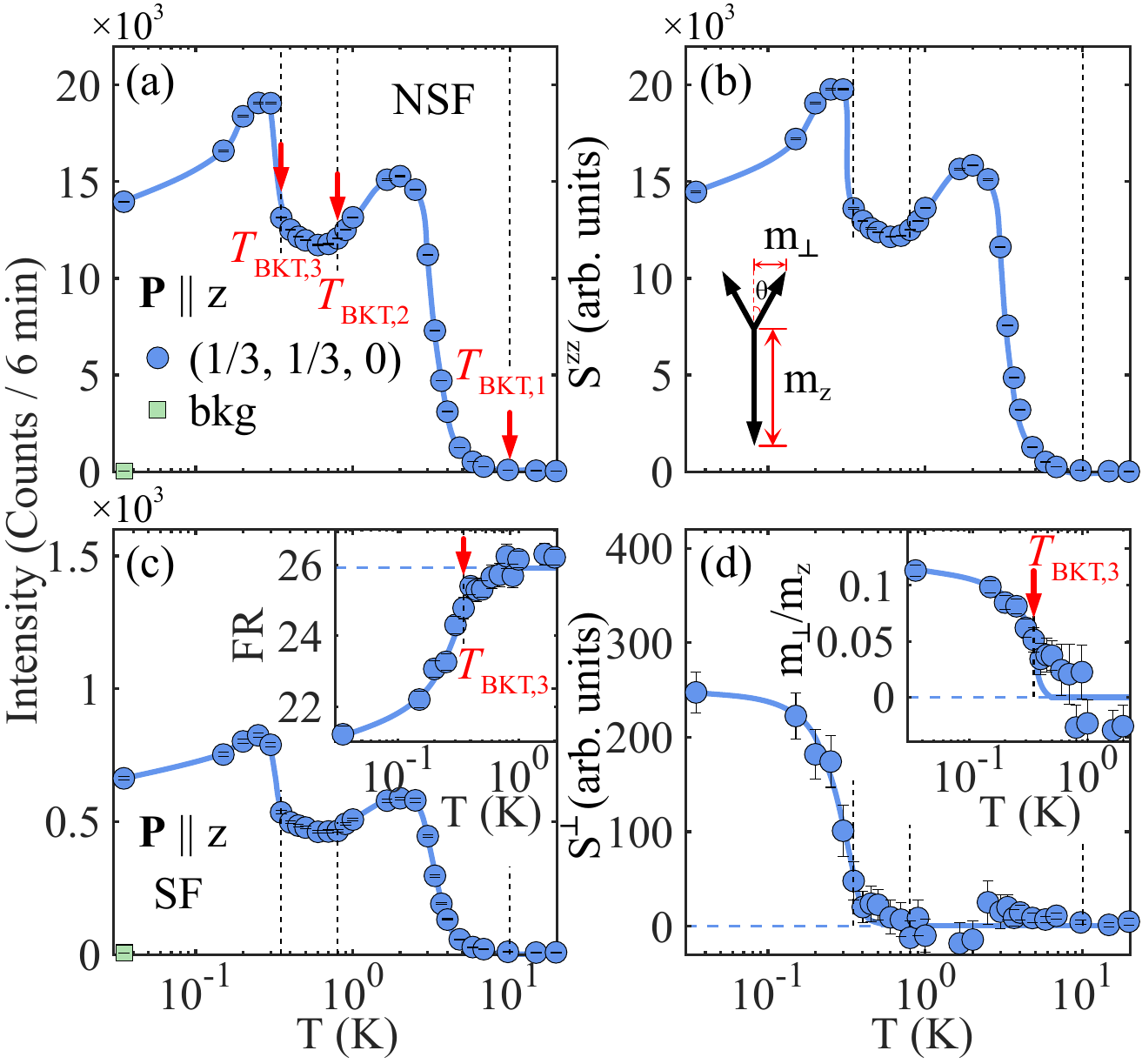}
	\caption{Temperature dependence of the magnetic intensity at $\mathbf{Q}=(1/3,1/3,0)$ measured using polarized neutron diffraction in the (a) non-spin-flip (NSF) and (c) spin-flip (SF) channels. Green squares denote the background. The incident neutron polarization is perpendicular to the scattering plane ($\mathbf{P}\parallel \mathbf{z}$). Red arrows and black dashed lines mark the BKT transitions. Inset in (c) shows the flipping ratio as a function of temperature. The pronounced decrease below $T_{\text{BKT,3}}$ signals the emergence of the transverse ordered moment. Blue dashed line indicates average flipping ratio between 0.5 and 2 K. (b),(d) Extracted longitudinal and transverse static structure factors as a function of temperature. Inset in (b) illustrates the quantum Y spin structure with zero net moment. Inset in (d) shows temperature dependence of the ratio of transverse and longitudinal ordered moments $m_{\perp}/m_z$. All temperature axes are in logarithmic scale. Solid lines are guides to the eye.  }\label{fig:polarized}
\end{figure}

Figure \ref{fig:polarized}(a) and \ref{fig:polarized}(c) show the temperature dependence of the NSF and SF intensities measured at $\mathbf{Q}=(1/3,1/3,0)$. Both channels reproduce the nonmonotonic behavior observed in the unpolarized neutron measurements [Fig.~\ref{fig:unpolarized_diffraction}(c)]. $I_{\text{BG,NSF}}$ and $I_{\text{BG,SF}}$ were estimated by rotating the sample by $\pm 7^\circ$ (green squares). If no transverse order were present ($S^{\perp}=0$), the flipping ratio $\mathrm{FR}=\frac{I_{\text{NSF}}-I_{\text{BG,NSF}}}{I_{\text{SF}}-I_{\text{BG,SF}}}$ would remain temperature independent. Instead, we observe a pronounced decrease at $T \approx 0.35$ K, coincident with $T_{\text{BKT,3}}$ \cite{Zhu2025}, as shown in the inset of Fig.~\ref{fig:polarized}(c). This provides direct evidence for the onset of a transverse ordered component. 

Using the average flipping ratio between 0.5 and 2 K [dashed line in Fig.~\ref{fig:polarized}(c) inset], we determine $\alpha = 0.9628(5)$ and extract the longitudinal and transverse static structure factors, plotted in Fig.~\ref{fig:polarized}(b) and \ref{fig:polarized}(d), respectively. The magnetic scattering is dominated by the longitudinal channel, consistent with the proximity of \KCSO to the Ising limit \cite{Zhu2024,Zhu2025,Chen2026}. While $S^{zz}$ develops already below $T_{\text{BKT,1}}$ and exhibits pronounced nonmonotonic temperature dependence, $S^{\perp}$ remains negligible until the lowest-temperature BKT transition at $T_{\text{BKT,3}}$, below which it increases continuously upon cooling.

Owing to the quasi-two-dimensional nature of the magnetic order, the static structure factor factorizes as $S(\mathbf{Q})\propto G(l)S_{\mathrm{2D}}(h,k)$, where $G(l)$ encodes interlayer correlations \cite{SM}. Consequently, the temperature evolution of $S^{zz}$ and $S^{\perp}$ reflects changes in both the ordered moments and interlayer correlations. However, the ratio $S^{\perp}/S^{zz}$ is largely insensitive to the common interlayer contribution therefore provides a robust measure of the relative magnitude of ordered moments $m_\perp/m_z$, as shown in the inset of Fig.~\ref{fig:polarized}(d). At the lowest measured temperature $T= 0.035$ K, we obtain $S^\perp/S^{zz}=|\langle\hat{m}_\perp(\mathbf{Q})\rangle/\langle\hat{m}_{z}(\mathbf{Q})\rangle|^2\approx0.017(2)$, where $\langle\hat{m}_{\perp/z}(\mathbf{Q})\rangle$ are the Fourier components of the transverse and longitudinal ordered moments. Given that the quantum Y structure in the supersolid phase exhibits no net moment, as illustrated in Fig.~\ref{fig:polarized}(b) inset \cite{kleine1992,gallegos2025}, and $|\langle\hat{m}_{\perp}(\mathbf{Q})\rangle|^2=3m_\perp^2$ and $|\langle\hat{m}_{z}(\mathbf{Q})\rangle|^2=\frac{9}{4}m_z^2$ \cite{SM}, we find that the transverse ordered moment amounts to only $\sim$11(1)\% of the longitudinal component. After correcting for the anisotropic $g$ factors ($g_{zz} = 7.9$; $g_{\perp} = 2.75$ \cite{SM}), this corresponds to a ratio of spin angular momenta $\langle\hat{S}_{\perp}\rangle/\langle\hat{S}_{z}\rangle \approx 32(2)\%$. 


We now perform a quantitative analysis of the interlayer correlations. The $l$ dependence of the magnetic scattering intensity measured with unpolarized neutrons is given by \cite{SM}
\begin{align}
    I(l)&\propto \sum\limits_{n,m} A_{nm} \exp\left(i\frac{2\pi l(n-m)}{3}\right) |F(\mathbf{Q})|^2 \nonumber\\
    &\left[\frac{1}{2}\left(1+\frac{\mathrm{Q}_z^2}{\mathrm{Q}^2}\right)S_{\mathrm{2D}}^{\perp}(h,k)+\left(1-\frac{\mathrm{Q}_z^2}{\mathrm{Q}^2}\right) S_{\mathrm{2D}}^{zz}(h,k)\right]\nonumber
\end{align}
where $A_{nm}=\langle \hat{\mathbf{S}}_n\hat{\mathbf{S}}_m\rangle/\langle \hat{\mathbf{S}}^2_n\rangle$ denotes the interlayer correlation, $\mathrm{Q}_z$ is the out-of-plane component of momentum transfer $\mathbf{Q}$, $\mathrm{Q}=|\mathbf{Q}|$, and $|F(\mathbf{Q})|^2$ is the squared magnetic form factor. 
A fit to the $l$-dependent scattering intensity at $\mathbf{Q}=(2/3,2/3,l)$ [solid lines in Fig.~\ref{fig:unpolarized_diffraction}(b),(d)] \cite{SM}, with $S_{\mathrm{2D}}^{\perp}/S_{\mathrm{2D}}^{zz}$ constrained to the values obtained independently from the polarized neutron measurements, reveals a striking change in the dominant interlayer correlations across $T_{\text{BKT,3}}$. At $T=0.5$ K, the leading correlation is between the AA-stacked layers separated by one lattice constant $c$ [Fig.~\ref{fig:unpolarized_diffraction}(b) inset] and antiferromagnetic, $A_{03}=-0.029(4)$, producing intensity maxima at half-integer $l$. In contrast, at $T=0.06$ K, the dominant correlation becomes ferromagnetic and between the second-nearest-neighbor layers separated by $\frac{2}{3}c$, $A_{02}=0.022(4)$, shifting the intensity maxima to $l=0$, $\pm$1.5 and $\pm$3.

\begin{figure}[th]
	\includegraphics[width=\columnwidth]{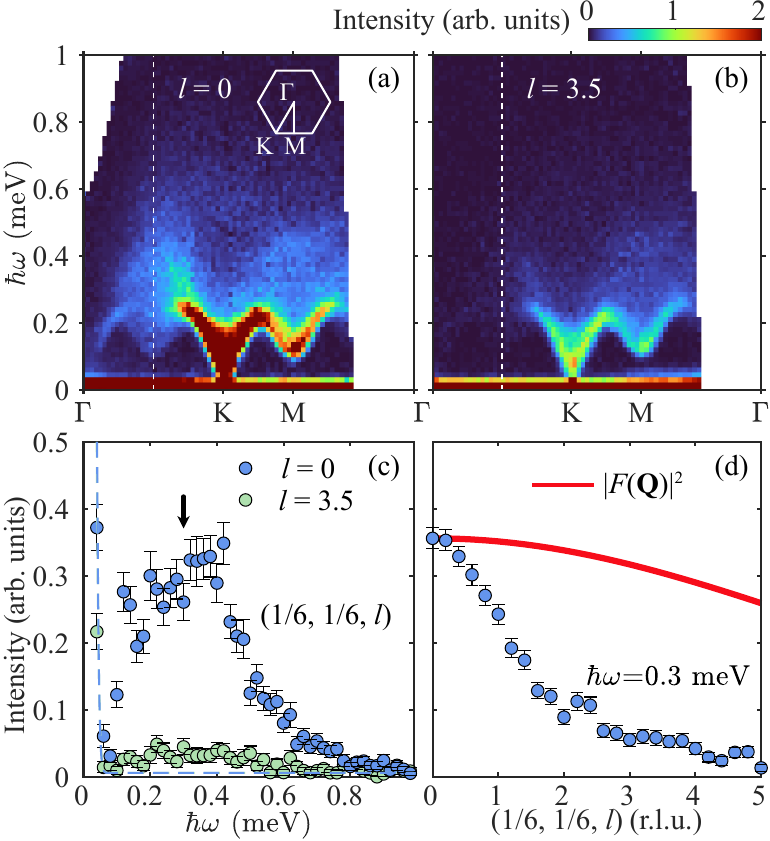}
	\caption{(a),(b) Magnetic excitation spectra of \KCSO measured by INS at $T$ = 0.1 K for $l=0$ and 3.5, respectively. White hexagon outlines the first Brillouin zone boundary. $\Gamma$, $K$ and $M$ denote high-symmetry positions. Dashed lines indicate the wave vector of the energy scans in (c). (c) Neutron scattering intensity as a function of energy at $\mathbf{Q}=(1/6,1/6,l)$ for $l=0$ and 3.5. Dashed  lines are fits to the elastic peak and constant background. Black arrow indicates the energy of the $l$ scan shown in (d). (d) Neutron scattering intensity as a function of $l$ at $\mathbf{Q}=(1/6,1/6,l)$ for $\hbar\omega=0.3$ meV. Red solid line shows the expected intensity drop due to the squared magnetic form factor alone. The suppression of spectral weight at large $l$ is substantially stronger than expected from the form factor. }\label{fig:INS_spectra}
\end{figure}

\begin{figure}
	\includegraphics[width=\columnwidth]{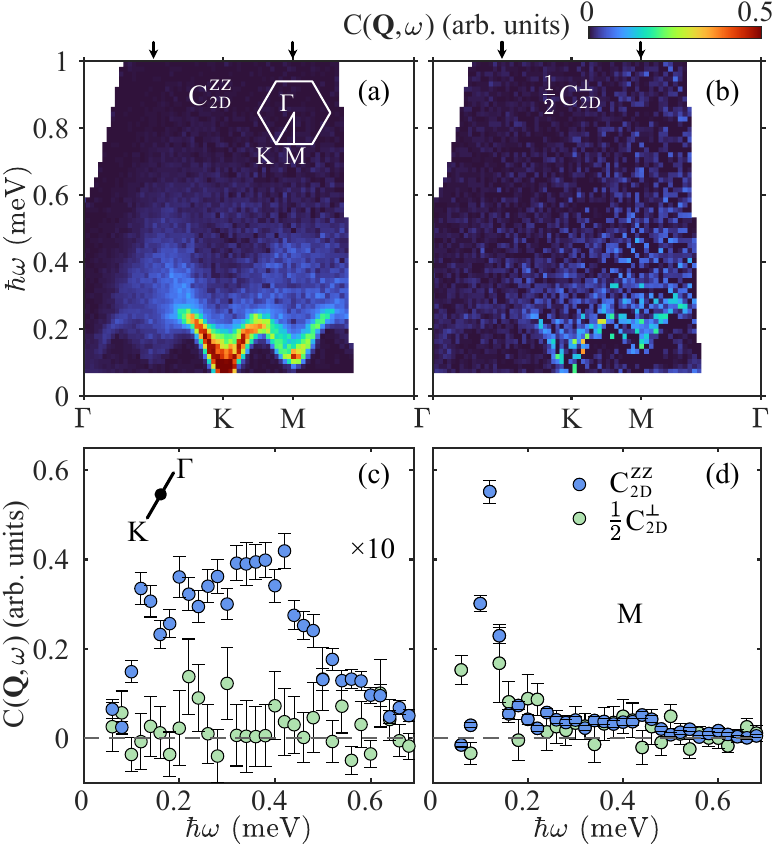}
	\caption{(a),(b) Extracted Fourier components of the longitudinal and transverse spin-spin correlation function $C^{zz}_\mathrm{2D}(\mathbf{Q},\omega)$ and $C^{\perp}_\mathrm{2D}(\mathbf{Q},\omega)$ in the supersolid phase, demonstrating the predominantly longitudinal character of the excitations. For the measured magnetic spectrum at $l=0$, $I\propto g_{zz}^2C^{zz}_\mathrm{2D} + g_\perp^2\frac{1}{2}C^{\perp}_\mathrm{2D}$. 
    Black arrows mark the wave vectors of the energy-dependent cuts in (c) and (d). (c),(d) Energy dependence of $C^{zz}_\mathrm{2D}$ and $C^{\perp}_\mathrm{2D}$ measured midway between $\Gamma$ and $K$ and at the $M$ point. Note the $\times$10 intensity scale factor.  }\label{fig:INS_polarization}
\end{figure}

Next, we investigate the polarization character of the low-energy spin fluctuations in the supersolid state. Because of the low energy scale and continuum nature of the excitations, polarized INS measurements are challenging. Instead, we use unpolarized neutrons and separate the longitudinal and transverse fluctuations by exploiting the polarization factor in the neutron-scattering cross section. 


INS measurements were performed on the AMATERAS cold-neutron time-of-flight spectrometer \cite{AMATERAS} at J-PARC using a single crystal oriented in the $(h,h,l)$ scattering plane and $E_i=2.24$ meV neutrons.
The magnetic excitation spectra measured at $l=0$ and $l=3.5$ 
are shown in Fig.~\ref{fig:INS_spectra}(a) and \ref{fig:INS_spectra}(b). For $l = 0$, the spectrum is consistent with previous studies \cite{Zhu2024,Zhu2025,Chen2026}, exhibiting a broad excitation continuum with roton-like minima at the $M$ point and midway between the $\Gamma$ and $K$ points. The overall spectrum is qualitatively similar for $l=3.5$, however, the spectral weight is strongly suppressed, particularly between $\Gamma$ and $K$ where the intensity nearly vanishes. A representative energy-dependent scan at $\mathbf{Q}=(1/6,1/6,l)$, indicated by white dashed lines in Figs.~\ref{fig:INS_spectra}(a) and ~\ref{fig:INS_spectra}(b), is shown in Fig.~\ref{fig:INS_spectra}(c). The observed suppression is substantially stronger than expected from the squared magnetic form factor alone, as illustrated in Fig.~\ref{fig:INS_spectra}(d) for $\hbar\omega=0.3$ meV (red solid line).

To elucidate this unusual behavior, we consider the neutron-scattering cross section. As the excitations are largely uncorrelated between triangular-lattice layers \cite{Chen2026}, the domain-averaged INS intensity can be written as
\begin{align}
   &I(\mathbf{Q},\omega) \propto |F(\mathbf{Q})|^2 \nonumber \\
   &\left[\frac{1}{2}\left(1+\frac{\mathrm{Q}_z^2}{\mathrm{Q}^2}\right) S_{\mathrm{2D}}^{\perp}(h,k,\omega)+\left(1-\frac{\mathrm{Q}_z^2}{\mathrm{Q}^2}\right)S_{\mathrm{2D}}^{zz}(h,k,\omega)\right] \nonumber
\end{align}
where $S^{zz/\perp}_{\mathrm{2D}}(h,k,\omega)$ are the longitudinal and transverse dynamical structure factors. 
Since neutrons probe only spin fluctuations perpendicular to the momentum transfer, the measured intensity at $l=0$ contains contributions from both channels, $I(l=0)\propto \frac{1}{2}S^\perp_{\mathrm{2D}}+S^{zz}_{\mathrm{2D}}$.
Increasing $l$ enhances the sensitivity to transverse fluctuations through the polarization factor $\left(1+\frac{\mathrm{Q}_z^2}{\mathrm{Q}^2}\right)$ while simultaneously suppressing the longitudinal contribution through $\left(1-\frac{\mathrm{Q}_z^2}{\mathrm{Q}^2}\right)$. If the excitations were predominantly transverse, the increase of the polarization factor with $l$ would partially compensate the intensity reduction due to the magnetic form factor. On the contrary, the experimentally observed loss of spectral weight at large $l$, which is substantially stronger than expected from the form factor alone [Fig.~\ref{fig:INS_spectra}(b)-(d)], provides direct evidence that the low-energy dynamics are dominated by longitudinal spin fluctuations.

This conclusion is corroborated by a quantitative decomposition of the measured spectrum. Figure~\ref{fig:INS_polarization}(a) and ~\ref{fig:INS_polarization}(b) show the extracted Fourier components of the longitudinal and transverse spin-spin correlation functions, $C^{\alpha\alpha}_{\mathrm{2D}}(\mathbf{Q}, \omega)=S^{\alpha\alpha}_{\mathrm{2D}}(\mathbf{Q}, \omega)/g_{\alpha\alpha}^2$. Representative energy-dependent cuts taken midway between $\Gamma$ and $K$ and at the $M$ point are presented in Fig.~\ref{fig:INS_polarization}(c) and ~\ref{fig:INS_polarization}(d). The results clearly demonstrate that the agreement between the longitudinal fluctuations calculated by QMC simulations and the INS spectra \cite{Zhu2025} cannot be attributed solely to the strongly anisotropic $g$ factor. Instead, the low-energy spin dynamics in the supersolid state are intrinsically dominated by the longitudinal channel across the measured momentum range, despite the small transverse ordered moment. 


Finally, we compare our results with theoretical predictions.
The observation of a small but finite transverse ordered moment in \KCSO demonstrates that the BEC order parameter survives in the ground state of the easy-axis XXZ model even in the near-Ising regime. This finding agrees with cluster mean field theory \cite{Yamamoto2014}, stochastic series-expansion calculations \cite{Zhang2011} and DMRG studies \cite{SellmannZhangEggert_PRB_2015_AnotherXXZtriangularPhD,gallegos2025}, but contrasts with exact diagonalization \cite{Ulaga2023,Ulaga2024} and tensor-network calculations \cite{Xu2025}, which predict a vanishing transverse ordered moment. Our measurements further enable a quantitative comparison with theory. Recent DMRG calculations \cite{gallegos2025} predict a transverse-to-longitudinal moment ratio of $\langle\hat{S}_{\perp}\rangle/\langle\hat{S}_{z}\rangle \approx 5\%$ for $J_{xy}/J_z = 0.07$, substantially smaller than the experimental value of $\approx32\%$. This discrepancy calls for a reexamination of the robustness of transverse order in the strong-Ising regime. Moreover, the transverse moment appearing precisely at the lowest-temperature BKT transition is in direct agreement with theoretical expectations \cite{Heidarian2018}.

The predominance of longitudinal low-energy spin fluctuations despite the small transverse order parameter is equally remarkable and sharply contrasts with the predictions of spin wave theory \cite{kleine1992,Kleine1992_Perturbationtheory,Zhu2024}. This behavior reflects the unconventional nature of the spin supersolid state. Unlike the classical Y state, the quantum supersolid carries no net ferromagnetic moment because of strong quantum fluctuations \cite{kleine1992,Kleine1992_Perturbationtheory,gallegos2025}. Its microscopic origin is best understood by starting from the strictly Ising limit, where the ground state is macroscopically degenerate with a residual entropy \cite{Wannier1950}. A finite $J_{xy}$ lifts this degeneracy through quantum fluctuations, selecting a long-range-ordered supersolid ground state and generating emergent low-energy excitations. As these excitations connect states within the same degenerate manifold of Ising ground states, they are predominantly longitudinal in character ($\Delta S_z=0$), in sharp contrast to conventional spin waves which are primarily transverse. By comparison, the transverse spin excitations in the supersolid phase originate from the spin-flip processes of the longitudinal ordered moments, therefore appear at much higher energies set by the Ising exchange $J_{zz} \approx 3.1$ meV \cite{Zhu2025,Chen2026}.

In summary, our neutron-scattering measurements on \KCSO demonstrate that the supersolid ground state persists in the near-Ising regime of the $S=1/2$ triangular-lattice XXZ model, characterized by a strongly suppressed transverse ordered moment coexisting with predominantly longitudinal low-energy spin dynamics. Together, these results reveal the highly unconventional nature of supersolidity driven by strong frustration and quantum fluctuations in quantum spin systems proximate to the Ising limit and, more broadly, in strongly interacting hard-core boson systems.

\begin{acknowledgements}
Work at ETHZ was supported by a MINT grant of the Swiss National Science Foundation. 
This work is based on experiments performed at the Swiss spallation neutron source SINQ, Paul Scherrer Institute, Villigen, Switzerland. Data at J-PARC were collected in Experiment no. 2025A0112. A portion of this research used resources at the Spallation Neutron Source, a DOE Office of Science User Facility operated by the Oak Ridge National Laboratory. The beam time was allocated to CORELLI on proposal number IPTS-34135. We acknowledge support from the Deutsche Forschungsgemeinschaft (DFG) through the W\"{u}rzburg-Dresden Cluster of Excellence on Complexity, Topology and Dynamics in Quantum Matter--$ctd.qmat$ (EXC 2147,
Project No.\ 390858490), as well as the support of the HLD at HZDR, member of the European Magnetic Field Laboratory (EMFL). The neutron scattering data collected at the ILL for the present work are available at Ref.~\cite{Zhu2024K2CoSeO3,Zhu2025ILL}. CRG-3155 beamtime on IN12 at the Institut Laue Langevin was supported by the Swiss State Secretariat for Education, Research and Innovation through a CRG grant.

\end{acknowledgements}

\bibliography{KCSO}

\end{document}


\title{Transverse order and longitudinal fluctuations in a near-Ising spin supersolid }
\author{Mengze Zhu}
\address{Laboratory for Solid State Physics, ETH Z\"{u}rich, 8093 Z\"{u}rich, Switzerland}

\author{V. Romerio}
\address{Laboratory for Solid State Physics, ETH Z\"{u}rich, 8093 Z\"{u}rich, Switzerland}

\author{S. Raymond}
\address{Université Grenoble Alpes, CEA, IRIG, MEM, MDN, F-38000 Grenoble, France}

\author{N. Murai}
\address{J-PARC Center, Japan Atomic Energy Agency, Tokai, Ibaraki 319-1195, Japan}

\author{S. Ohira-Kawamura}	
\address{J-PARC Center, Japan Atomic Energy Agency, Tokai, Ibaraki 319-1195, Japan}

\author{K.~Yu.~Povarov}
\address{Dresden High Magnetic Field Laboratory (HLD-EMFL) and W\"urzburg-Dresden Cluster of Excellence ctd.qmat, Helmholtz-Zentrum Dresden-Rossendorf, 01328 Dresden, Germany}

\author{S.~A.~Zvyagin}
\address{Dresden High Magnetic Field Laboratory (HLD-EMFL) and W\"urzburg-Dresden Cluster of Excellence ctd.qmat, Helmholtz-Zentrum Dresden-Rossendorf, 01328 Dresden, Germany}

\author{R. Sibille}
\address{PSI Center for Neutron and Muon Sciences, 5232 Villigen PSI, Switzerland}

\author{A. Minelli}
\address{Neutron Scattering Division, Oak Ridge National Laboratory, Oak Ridge, Tennessee 37831, USA}

\author{Z. Yan}
\address{Laboratory for Solid State Physics, ETH Z\"{u}rich, 8093 Z\"{u}rich, Switzerland}

\author{S. Gvasaliya}
\address{Laboratory for Solid State Physics, ETH Z\"{u}rich, 8093 Z\"{u}rich, Switzerland}

\author{A.~Zheludev}
\email{zhelud@ethz.ch; http://www.neutron.ethz.ch/}
\address{Laboratory for Solid State Physics, ETH Z\"{u}rich, 8093 Z\"{u}rich, Switzerland}

\begin{abstract}
\end{abstract}

\date{\today}
\maketitle

\section{Fitting the $l$ dependence of the unpolarized neutron diffraction data}
\subsection{Unpolarized neutron scattering cross section}
The elastic magnetic scattering intensity for unpolarized neutrons is given by \cite{Squiresbook}
\begin{align}
    I(\mathbf{Q}) = & \lim\limits_{t\rightarrow\infty}(\frac{\gamma r_0}{2})^2 |F(\mathbf{Q})|^2 \exp(-2W)\sum\limits_{\alpha\beta}(\delta_{\alpha\beta}-\frac{Q_{\alpha} Q_{\beta}}{Q^2})\nonumber \\
    &\times g_\alpha g_\beta\sum\limits_{ij}\exp(i\mathbf{Q}(\mathbf{r}_i-\mathbf{r}_j)) \langle \hat{S}_i^\alpha(0) \hat{S}_j^\beta(t)\rangle, \nonumber
\end{align}
where $\gamma r_0=0.54\times 10^{-12}$ cm, $|F(\mathbf{Q})|^2$ is the squared magnetic form factor, and $\exp(-2W)$ is the Debye-Waller factor.

We label each spin by $i\rightarrow (n,\mu)$, where $n$ denotes the layer index and $\mu$ labels a magnetic site within a triangular-lattice plane. Assuming that every layer hosts the same in-plane magnetic structure, $\mathbf{r}_{n\mu}=\mathbf{R}_\mu+z_n \hat{c}$ and $\hat{S}_{n\mu}^{\alpha}=\eta_n\hat{S}_\mu^\alpha$, where $\eta_{n}$ describes the relative spin orientation between layers. The spin correlation function then factorizes as $\langle \hat{S}_{n\mu}^\alpha(0) \hat{S}_{m\nu}^\beta(t)\rangle = \langle \eta_n\eta_m\rangle \langle S_\mu^\alpha(0) S_\nu^\beta(t) \rangle$. 

For the ABC stacking of triangular-lattice layers separated by $c/3$, the interlayer contribution becomes
\begin{align}
   \langle \eta_n \eta_m\rangle \exp(i\mathbf{Q}(z_n-z_m)\hat{c}) = A_{nm}\exp\left(i\frac{2\pi l(n-m)}{3}\right), \nonumber 
\end{align}
with $A_{nm}\equiv\langle\eta_n\eta_m\rangle$ parametrizes the interlayer spin correlations. For perfectly ordered ferromagnetic and antiferromagnetic stacking, $A_{nm}=1$ and $-1$, respectively.

The contribution from the spin correlation within the triangular plane is
\begin{align}
\sum\limits_{\alpha\beta}(\delta_{\alpha\beta}-\frac{Q_{\alpha} Q_{\beta}}{Q^2})\nonumber
    \times g_\alpha g_\beta\sum\limits_{\mu\nu} \langle S_\mu^\alpha S_\nu^\beta \rangle \exp(i\mathbf{Q}(\mathbf{R}_\mu-\mathbf{R}_\nu)). \nonumber
\end{align}

For the collinear phase, where the ordered moments are along the $z$ axis, only the longitudinal spin component contributes,
\begin{align}
    (1-\frac{Q_z^2}{Q^2})S^{zz}_{\mathrm{2D}}(h,k), \nonumber
\end{align}
with 
\begin{align}
  S^{\alpha\alpha}_{\mathrm{2D}}(h,k)=&N_m\frac{(2\pi)^2}{A_{0m}}  \sum\limits_{\mathbf{\tau}_m}\delta(\mathbf{Q}-\bm{\tau}_m) \nonumber\\&|g_{\alpha\alpha}\langle S^\alpha\rangle\sum\limits_{d}\sigma_d\exp(i\bm{\tau}_m \mathbf{d})|^2, \nonumber
\end{align}
where $N_m$ and $A_{0m}$ are the number and area of the 2D magnetic unit cells, $\bm{\tau}_m$ is a magnetic reciprocal lattice vector, $\bm{d}$ runs over the basis sites in the magnetic unit cell, and $\langle S^\alpha\rangle $ denotes the averaged over one magnetic sublattice.

The supersolid Y phase possesses a continuous U(1) degeneracy corresponding to rotations of the transverse spin component. Since a macroscopic crystal consists of domains with all possible orientations, averaging over the degenerate domains yields
\begin{align}
    \frac{1}{2}(1+\frac{Q_z^2}{Q^2})S^{\perp}_{\mathrm{2D}}(h,k) +(1-\frac{Q_z^2}{Q^2})S^{zz}_{\mathrm{2D}}(h,k), \nonumber
\end{align}
where $S^\perp_{\mathrm{2D}} = S^{xx}_{\mathrm{2D}}+S^{yy}_{\mathrm{2D}}$.

Combining the in-plane and interlayer contributions gives the $l$ dependence of the magnetic scattering intensity,
\begin{align}
    I(l)&\propto \sum\limits_{nm} A_{nm} \exp\left(i\frac{2\pi l(n-m)}{3}\right) |F(\mathbf{Q})|^2\nonumber\\ 
   &\left[\frac{1}{2}\left(1+\frac{\mathrm{Q}_z^2}{\mathrm{Q}^2}\right) S^{\perp}_{\mathrm{2D}}(h,k)+\left(1-\frac{\mathrm{Q}_z^2}{\mathrm{Q}^2}\right) S^{zz}_{\mathrm{2D}}(h,k)\right]. \nonumber
\end{align}

\subsection{Fit results for interlayer correlations}

Elastic neutron diffuse scattering maps were acquired on CORELLI using the cross-correlation chopper. Reciprocal-space maps in the $(h,h,l)$ plane were collected by rotating the sample through 300$^\circ$ about the vertical axis in 3$^\circ$ increments, with a counting time of 3 min per step. Data reduction and reciprocal-space reconstruction were carried out using Mantid.

The interlayer correlations were determined by fitting the $l$ dependence of the magnetic scattering intensity measured at $\mathbf{Q} = (2/3,2/3,l)$ for each temperature, with $S^\perp_{\mathrm{2D}}/S^{zz}_{\mathrm{2D}}=0$ and 0.017 at $T$ = 0.5 K and 0.06 K, respectively, as determined independently from the polarized neutron measurements, and a constant background. Equivalent scans measured at $(-2/3,-2/3,l)$ were averaged to improve counting statistics.
Since the interlayer correlations are weak, the fits were performed by
retaining only correlations up to the third-nearest-neighbor layer,
\begin{align}
    &\sum\limits_{nm} A_{nm} \exp\left(\mathrm{i}\frac{2\pi l(n-m)}{3}\right) \approx \nonumber \\ &1+2A_{01}\cos(\frac{2\pi l}{3}) + 2A_{02}\cos(\frac{4\pi l}{3})+2A_{03}\cos(2\pi l). \nonumber
\end{align}

At $T=0.5$ K, the fit yields $A_{01}=0.008(4)$, $A_{02}=0.005(4)$, $A_{03}=-0.029(4)$ (red solid lines in Fig.~1(d) of the main text). The negative value of $A_{03}$ indicates antiferromagnetic correlations between the third-nearest-neighbor (AA stacked) planes, giving rise to intensity maxima at half-integer $l$. 

At $T=0.06$ K, the fit yields $A_{01}=0.003(4)$, $A_{02}=0.022(4)$, $A_{03}=0.004(4)$ (red solid lines in Fig.~1(b) of the main text), indicating predominantly ferromagnetic correlations between the second-nearest-neighbor layers, giving rise to intensity maxima at $l$ = 0, $\pm$1.5, and $\pm$3.

\subsection{Consistency check}
\begin{figure}[ht]
	\includegraphics[width=\columnwidth]{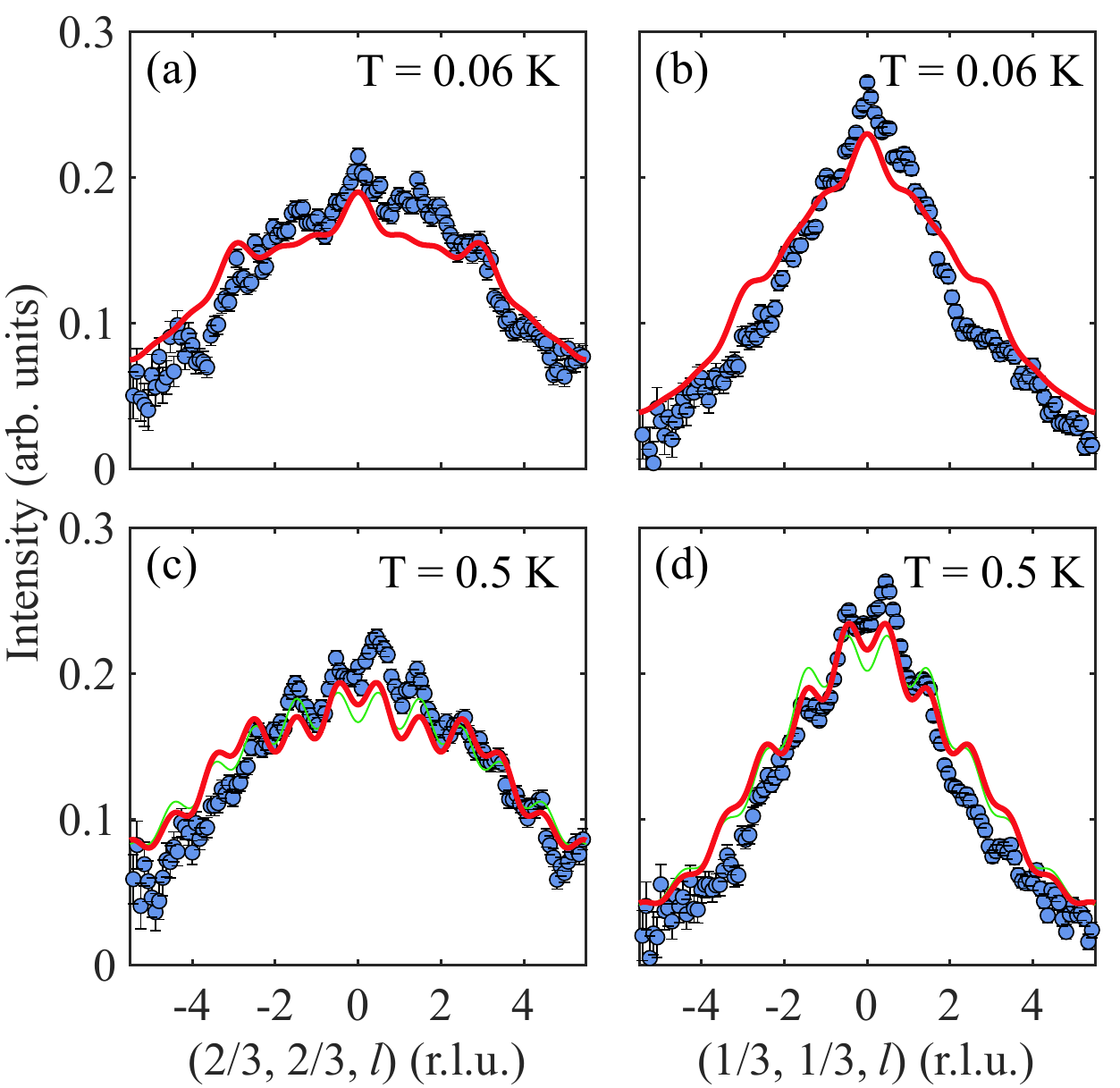}
	\caption{Background-subtracted $l$ scans at $\mathbf{Q} = (2/3,2/3,l)$ and $(1/3,1/3,l)$ measured at (a),(b) $T=0.06$ K and (c),(d) 0.5 K, respectively. Solid lines are global fits described in the text. }\label{fig:lscans_fit}
\end{figure}

As a consistence check, we also performed a global fit to the $l$ dependence of the magnetic scattering intensity measured at both $\mathbf{Q}=(1/3,1/3,l)$ and $(2/3,2/3,l)$. However, the $(1/3,1/3,l)$ reflection is strongly affected by the background scattering from the direct beam [see Fig.~1(a) of the main text]. Fig.~\ref{fig:lscans_fit} shows the background-subtracted $l$ scans measured at $\mathbf{Q} = (1/3,1/3,l)$ and $(2/3,2/3,l)$ at $T$ = 0.06 K and 0.5 K.
The data were integrated over $[H,H]=[0.3,0.36]$ and $[0.63,0.69]$, respectively, and over $[H,-H]=[-0.05,0.05]$ in the transverse directions.
The background was estimated from regions slightly off the magnetic reflection positions, namely $[H,H]=[0.23,0.26]$ and $[0.4,0.43]$ for the (1/3,1/3,0) reflection, and $[H,H]=[0.56,0.59]$ and $[0.73,0.76]$ for the (2/3,2/3,0) reflection. As before, equivalent scans measured at $(-1/3,-1/3,l)$ and $(-2/3,-2/3,l)$ were averaged to improve the counting statistics.

At $T=0.5$ K, the global fit yields $A_{01}=0.035(6)$, $A_{02}=0.004(6)$, $A_{03}=-0.033(6)$ with $\chi^2=15.6$ (red solid lines in Fig.~\ref{fig:lscans_fit}). The negative value of $A_{03}$ is consistent with the analysis based sorely on the $(2/3,2/3,l)$ reflection, confirming antiferromagnetic correlations between third-nearest-neighbor layers. The positive value of $A_{01}$ suggests weak ferromagnetic correlations between adjacent layers, producing intensity maxima at $l=0$ and 3. However, $A_{01}$ is sensitive to the background subtraction.  A fit with $A_{01}=0$, $A_{02}=0.004(6)$, $A_{03}=-0.033(6)$, yields a comparable goodness of fit, $\chi^2=18.1$ (green solid lines), indicating that the nearest-neighbor interlayer correlation cannot be uniquely determined by the present data. 

At $T=0.06$ K, the global fit yields $A_{01}=0.036(6)$, $A_{02}=0.021(7)$, $A_{03}=0.011(7)$ (red solid lines). The positive $A_{02}$ is likewise consistent with the analysis of the $(2/3,2/3,l)$ reflection, confirming the ferromagnetic correlations between the second-nearest-neighbor layers. Although the global fit also yields a positive value of $A_{01}$, this parameter is primarily determined by the scattering intensity near $l=0$ and is therefore particularly sensitive to the background subtraction of the $(1/3,1/3,l)$ reflection.

\section{Extracting $g_\perp$ from ESR measurements}
\begin{figure}[ht]
	\includegraphics[width=\columnwidth]{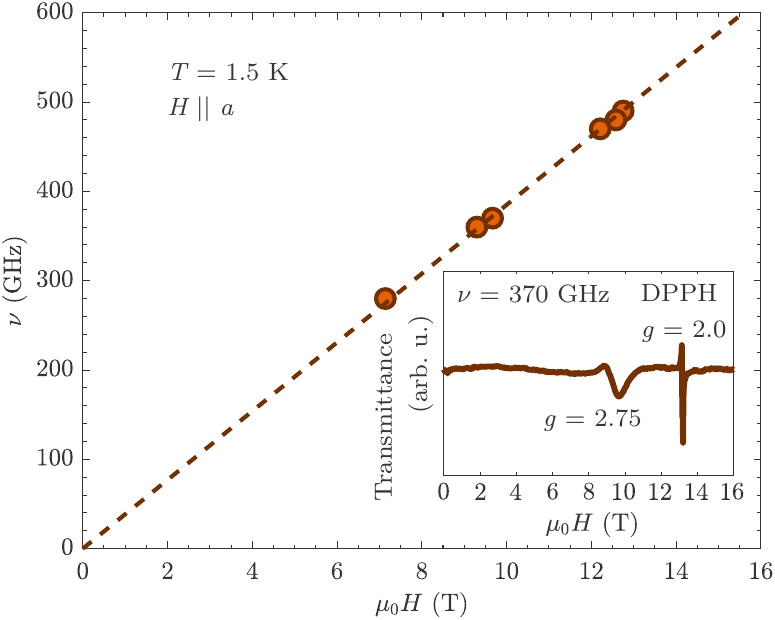}
	\caption{Frequency-field diagram of low-energy magnetic excitations in K$_2$Co(SeO$_3$)$_2$, $H \parallel a$. Points are the resonance-line centers; the dashed line corresponds to $g_\perp=2.75$. The inset shows an exemplary resonance line at $\nu=370$~GHz; sharp resonance on the right is the DPPH marker.  }\label{fig:ESR}
\end{figure}

The transverse $g$ factor, $g_\perp$, was determined from electron spin resonance (ESR) measurements. For the ESR experiments, we employed a spectrometer with a sample holder in the Voigt configuration (similar to the one described in Ref.~\cite{Zvyagin_PhysB_2004_ESRinCuGeO3}), equipped with a $16$~T superconducting magnet. We used VDI microwave-chain sources (product of Virginia Diodes, Inc., USA) to generate radiation in the frequency range of c.a. $200-500$~GHz, and a hot-electron n-InSb bolometer (product of QMC Instruments Ltd., UK), operated at $4.2$~K,  as a THz detector.  We used 2,2-diphenyl-1-picrylhydrazyl (DPPH) as a standard frequency-field marker. The results are displayed in Fig.~\ref{fig:ESR}. Low-temperature ESR signal for $H\parallel a$ magnetic field orientation can be well described by the linear frequency-field dependence with $g_\perp=2.75$.

\bibliography{KCSO}